\newcommand{\thedate}{20-03-2020}
\newcommand{\edition}{ed01}
\newcommand{\version}{A completely revised arXiv:\ 1911.04148 [hep-ph],\ \ \thedate\ (\edition)}

\documentclass[%
onecolumn,%
oneside,%
floats,%
aps,%
prd,%
nobibnotes,%
nofootinbib,%
amsmath,%
amssymb,%
amsfonts,%
superscriptaddress,%
11pt
]{revtex4}

\usepackage[utf8]{inputenc}
\usepackage{graphicx,array,dcolumn}
\usepackage{cases}
\usepackage[paperwidth=210mm,paperheight=297mm,centering,hmargin=1.8cm,vmargin=2.5cm]{geometry}
\usepackage{enumerate}

\usepackage{hyperref}

\usepackage[normalem]{ulem}
\usepackage{soul}
\usepackage{bm}
\usepackage{bold-extra}

\def\fun#1#2{\lower3.6pt\vbox{\baselineskip0pt\lineskip.9pt
\ialign{$\mathsurround=0pt#1\hfil ##\hfil$\crcr#2\crcr\sim\crcr}}}

\newcommand{{\SD}}{\rm SD}

\newcommand{{\Mc}}{\mathcal{M}}

\newcommand{\lb}{\left(}
\newcommand{\rb}{\right)}

\def\-g{\sqrt{-g}}

\newcommand{\be}{\begin{equation}}
\newcommand{\ee}{\end{equation}}

\newcommand{\ben}{\begin{equation*}}
\newcommand{\een}{\end{equation*}}

\newcommand{\bea}{\begin{eqnarray}}
\newcommand{\eea}{\end{eqnarray}}

\renewcommand\rho{\varrho}

\newcommand\Rot{\operatorname{rot}}
\newcommand\Div{\operatorname{div}}

\begin{document}


\title{\sc \Large{the rayleigh bubble in quark matter under a strong magnetic field}}

\thanks{\version} 


\author{\firstname{B.O.}~\surname{Kerbikov}\medskip}

\email{borisk@itep.ru}

\affiliation{NRC ``Kurchatov Institute'' – Institute for Theoretical and Experimental Physics,
Moscow 117218, Russia \smallskip}

\affiliation{Lebedev Physical Institute, Moscow 119991, Russia \smallskip}

\affiliation{Moscow Institute of Physics and Technology, Dolgoprudny 141700,
Moscow Region, Russia \bigskip}

\author{\firstname{M.S.}~\surname{Lukashov}\medskip}

\email{m.s.lukashov@gmail.com}

\affiliation{NRC ``Kurchatov Institute'' – Institute for Theoretical and Experimental Physics,
Moscow 117218, Russia \smallskip}

\date{\today}

\begin{abstract}

\noindent The transition of QGP from fluid-dynamical regime to freeze-out is accompanied by the onset of instabilities. In the present paper we investigate the impact of the magnetic field on the Rayleigh instability. We show that extremely strong field generated in peripheral heavy ion collisions has an insignificant influence on the Rayleigh bubble dynamics. Magnetic ``friction'' turns out to be much weaker than the viscous one.
 

\end{abstract}

\maketitle

\large

\section{Introduction \label{intro} }

The evolution of the fireball created in heavy ion collisions is at the early stages adequately described by hydrodynamics \cite{01,02,03,04}. At a certain phase hydrodynamics breaks down but when and how it happens remains to a great extent ill understood \cite{05,06,07,08,09}. Numerous attempts have been performed in recent years to clarify the character of the transition from the hydrodynamic regime to the chemical freezeout stage \cite{07,08,09,10,11,12}. Different kinds of instabilities accompanying this transition have been predicted \cite{07,08,09,13,14,15,16,17}. Here we focus on one particular instability, namely cavitation \cite{08,09,18,19,20,21} and closely related Rayleigh collapse \cite{22,23,24,25,26}. To our knowledge the possibility of the Rayleigh collapse and sonoluminescence in quark matter were first discussed in \cite{21}. Cavitation is a phenomenon in which rapid change of pressure in a liquid leads to the formation of small bubbles where the pressure is relatively low. The mechanism responsible for the formation of low and even negative pressure has been proposed in \cite{08,09,20}. The driving force leading to the negative effective pressure and to the onset of cavitation is the enchanced bulk viscocity. Enchanced bulk viscosity also leads to the anomalous sound attenuation due to Mandelshtam-Leontovich slow mode formation \cite{28,29}. The theory of bubble dynamics and the bubble collapse were started by Lord Rayleigh in $1917$ \cite{22} when he investigated cavitation damage of ship propellers. The Rayleigh equation describing the collapse reads 
\be
R\,\ddot{R} + \dfrac32\,\dot{R}^{2} = 0, 
\label{eq01}
\ee 
where $R(t)$ is the bubble radius. The solution of (\ref{eq01}) is the power law
\be
R(t) \thicksim (t_{C} - t)^{2/5}, 
\label{eq02}
\ee
which leads a divergent wall velocity $\dot{R}(t) \sim (t_{C} - t)^{-3/5}$ at $t \to t_{C}$. At collapse the interior of the bubble gets compressed, heats up, emits a shock sound wave and possibly emits light. In the next section we shall derive the collapse singularity on purely dimensional grounds. Accordingly to the present understanding \cite{23,24,25,26,27} Eq.(\ref{eq01}) gives an oversimplified picture of the collapse with only inertia forces  accounted for. Depending on what additional factors and parameters are included into consideration and what simplifying assumptions are made the equation of the bubble motion and pulsation takes different forms. The simplest version of the celebrated Rayleight-Plesset (R-P) equation reads \cite{23,27} 
\be
R\,\ddot{R} + \dfrac32\,\dot{R}^{2} = \dfrac{1}{\rho}\left( p_L - p_0 - P(t)\right), 
\label{eq03}
\ee
where $\varrho$ is the liquid density assumed to be a constant, $p_L$ the pressure in the liquid at the bubble wall, $p_0$ the ambient pressure, $P(t)$ the driving acoustic pressure. 
Equation (\ref{eq03}) corresponds to the incompressible liquid. For the compressible liquid acoustic corrections have been considered in \cite{30}. Viscous losses and the surface tension will be included in the next section. The review of a large number of publications on different aspects of the bubble dynamics is beyond the scope of this work. 

The quark matter formed in heavy ion collisions is subjected to a strong magnetic field generated by spectator protons and this field may be captured by a fireball provided it has a finite electric conductivity. The number of publications on the magnetic field related phenomena in QGP is overwhelming, see, e.g. \cite{31,32} and references therein.
The main message is that the intense magnetic field has a strong, sometimes drastic, influence on the QCD phase diagram \cite{33}, thermodynamic properties \cite{34}, meson masses \cite{35}, transport coefficients \cite{36,37}. This list can be continued. Necessary to emphasize that the above mentioned numerous calculations have been performed taking the magnetic field to be time-independent and uniform. The solution of a realistic problem with the magnetic field and the fireball geometry changing in time remains a task for future work. The goal of this paper is to investigate the influence of the magnetic field on the bubble dynamics and collapse in quark matter. The somewhat unexpected conclusion will be that strong magnetic field has a negligible impact on the bubble dynamics. The shear viscosity turns out to be much more important despite the well-known statement that the QGP is the most ideal fluid in nature. 

The picture of the Rayleigh bubble evolution is complicated. Different authors (see below) include various factors into consideration. Our aim is to get a clear-cut conclusion on the role of the magnetic field in the Rayleigh instability. In this context we shall resort to several approximations specified in what follows.


The paper is organized as follows. In Sect.II we derive the Rayleigh collapse solution on purely dimensional grounds. In Sect.III the systematic derivation of the R-P equation with viscous losses and surface tension is presented. Section IV is the core of the paper. Here we obtain the expression for the magnetic force acting on the bubble and include this term into the R-P equation. In Section V we show that the effect of an extremely strong magnetic field is very small and explain the reason for that. In concluding Section VI we discuss the possible composition of matter inside and outside the bubble. 


\section{Derivation of the Rayleigh Collapse on Dimensional Grounds}

The R-P collapse described by Eq.(\ref{eq03}) and its more involved forms is an intricated physical phenomenon depending on multiple sets of parameters and boundary conditions. Following \cite{38} we shall show that a singular solution may be found simply on dimensional grounds provided the important parameters are correctly chosen. The time development of the collapse depends on the wall radial coordinate $R$, the wall radial velocity $\dot{R}$, density $\varrho$, pressure $p$, and temperature $T$. At the initial moment the system is in equilibrium with $\dot{R}=0$, $\varrho=\varrho_0$, $p=p_0$, and $T=T_0$. Another important quantity is the energy $E_0$ inserted into the system. The energy $E_0$ may be considered as the work done on the bubble by the pressure which would exist at the location of the center of the bubble were the bubble not to be present. The dimensional formula for the bubble radius as a function of time is defined as the expression in terms of the basic parameters $\varrho_0$, $p_0$, $E_0$ with proper dimensions. The dimension of $E_0$ deserves a comment. The work done by the pressure results in the change of the liquid kinetic energy. The kinetic energy has a dimension $MR^2/t^2$. We shall return to this point after Eq.(\ref{eq13}). On physical grounds the number of essential parameters may be reduced from three to two. For strong collapse the nonlinearity induced by the shock wave from the bubble is much greater than the nonlinearity in the initial wave pressure. Therefore the value of the initial pressure becomes inessential. This is equivalent to the original Rayleigh assumption expressed by Eq.(\ref{eq01}) that only liquid inertia mattered. In this way we arrive at the following dimensional equation 
\be
R^{\delta}=\rho_0^{\nu}E_0^{\mu}t^{\xi}=(MR^{-3})^{\nu}(MR^2/t^2)^{\mu}t^{\xi}. 
\label{eq04}
\ee
This yields a system of equations 
\bea
-3\nu + 2\mu &=& \delta,\nonumber\\
\nu + \mu &=& 0,\label{eq05}\\
-2\mu + \xi &=& 0.\nonumber
\eea
We have three equations for four parameters so that one parameter remains arbitrary. To proceed we form the time/radius relation $\xi/\delta$. Then from (\ref{eq05}) one easily gets $\xi/\delta = 2/5$, so that 
\be
R = A t^{2/5} 
\label{eq06}
\ee
in a complete agreement with the Rayleigh result (\ref{eq02}). Formula (\ref{eq06}) may be looked at as a scaling law.

\section{The Inclusion of Viscous Losses and Surface Tension}

The aim of this work is to include the magnetic force into the R-P equation and to compare its role to that of the viscous forces. As a first step we shall explain how to incorporate the viscous and the surface tension terms into the R-P equation. For the detailed derivation and the description of the necessary assumptions we refer the reader to \cite{24,26,30}.

The starting point is the Navier-Stokes equation for the incompressible liquid which we will write following \cite{39} 
\be
\dfrac{\partial {\bf v}}{\partial t} + ({\bf v}{\bf {\stackrel{\rightarrow}{\nabla}}}){\bf v} = -\dfrac{1}{\rho}\,{\bf {\stackrel{\rightarrow}{\nabla}}}p + \dfrac{\eta}{\rho}\,\Delta{\bf v} + \dfrac{1}{\rho}\,{\bf f},
\label{eq07}
\ee
where $\eta$ is the shear viscosity and ${\bf f}$ refer to the external forces. In our case $\bf{f}$ is the magnetic force which will be considered in the next section. 

Alternatively, on more general grounds one can start from the relativistic version of the Euler equation \cite{40}

\be
\dfrac{\partial {\bf v}}{\partial t} + ({\bf v}{\bf {\stackrel{\rightarrow}{\nabla}}}){\bf v} = - \dfrac{1}{(e + p)\gamma^2}\left[ {\bf {\stackrel{\rightarrow}{\nabla}}}p + {\bf v} \dfrac{\partial {p}}{\partial t} \right],
\label{eq08*}
\ee

where $e$ is the total energy density, $\gamma$ is the Lorentz factor. We do not follow this approach for two reasons. First, the addition of viscosity amounts to additional non-diagonal terms that makes the equation very complicated \cite{40,41}. Second, as it was shown in \cite{42} the small difference between non-viscous relativistic and non-relativistic cases arises at the last stages of the bubble collapse. However, other factors like sono-luminescence \cite{25,43} are important at the end of the collapse. Embedding the Rayleigh collapse into the Bjorken flow \cite{44} is a challenging and difficult problem to be solved in future. Based on the results presented below one can safely ignore the magnetic field in solving the relativistic or Bjorken-like Rayleigh collapse.  

The  flow in (\ref{eq07}) is redial and therefore the viscous term in (\ref{eq07}) is not important. We shall see shortly that viscosity enters through the pressure difference inside and outside the wall. For the spherical bubble with the radius $R(t)$ and purely radial wall velocity (\ref{eq07}) takes the form
\be
\dfrac{\partial v}{\partial t} + v\,\dfrac{\partial v}{\partial r} + \dfrac{1}{\rho}\,\dfrac{\partial p}{\partial r} = 0.
\label{eq09}
\ee
For the radial and irrotational motion the velocity can be represented in terms of a potential $\varphi$ as $\bf{v} = {\stackrel{\rightarrow}{\nabla}}\varphi$, or $v = \partial\varphi/\partial{r}$. Then 
\be
v(r,t)=\dfrac{R^2(t)}{r^2(t)}\dot{R}(t), \quad \phi(r,t) = -\dfrac{R^2(t)}{r(t)}\dot{R}(t).
\label{eq10}
\ee
In terms of $\varphi$ (\ref{eq09}) takes the form
\be
\dfrac{\partial^2 \varphi}{\partial t \partial r} + \dfrac12\dfrac{\partial}{\partial r}\lb \dfrac{\partial \phi}{\partial r} \rb^2 + \dfrac{1}{\rho}\,\dfrac{\partial p}{\partial r} = 0.
\label{eq11}
\ee
Integrating over $r$ we arrive at
\be
\dfrac{\partial\varphi}{\partial t} + \dfrac12\lb \dfrac{\partial \phi}{\partial r} \rb^2 = -\dfrac{p(r)-p_0}{\rho},
\label{eq12}
\ee
where $p_0$ is the static pressure outside the bubble wall. Using expression (\ref{eq10}) for $\varphi$ and evaluating (\ref{eq12}) at $r=R$ we obtain
\be
R\,\ddot{R} + \dfrac32\,\dot{R}^2 = \dfrac{p_L-p_0}{\rho},
\label{eq13}
\ee
where $p_L$ is the liquid pressure at the outer interface of the bubble.

Expression (\ref{eq10}) for the bubble expansion or contraction velocity allows to calculate the kinetic energy entering into the dimensional equation (\ref{eq04}). Consider the evolution of the liquid spherical shell with the surface area $4 \pi r^2$ and the thickness $dr$.  The corresponding kinetic energy is 
\be
E = 4\pi\rho \int_{R}^{\infty} dr r^2 \left[ \dfrac{R^2(t)}{r^2} \dot{R}(t) \right]^2 = 4 \pi \rho R^3 \dot{R}^2 \sim M\dfrac{R^2}{t^2}
\label{eq14}
\ee
The confirms the choice of the energy dimension in (\ref{eq04}).

Next we have to include the surface tension and the viscous correction. To this end we consider the pressure difference between the outside value $p_L$ and the inside one $p_i$. We use the Laplace-Young expression for the surface tension \cite{41} and assume that the fluid is a Newtonian one so that the Couchy viscous stress is linearly proportional to the radial strain $\partial v / \partial r$ \cite{41}. The matching condition reads
\be
p_i - \lb p_L - 2\,\eta\,\dfrac{\partial v}{\partial r}\rb = p_i - p_L - 4\,\eta\,\frac{\dot{R}}{R} = \dfrac{2\sigma}{R},
\label{eq15}
\ee
or
\be
p_L= p_i - \dfrac{2\sigma}{R} - 4\,\eta\,\frac{\dot{R}}{R}.
\label{eq16}
\ee
Substitution this into (\ref{eq13}) leads \cite{25}
\be
R\,\ddot{R} + \dfrac32\,{\dot{R}}^2 = \dfrac{1}{\rho}\lb p_i - p_0 - \dfrac{2\sigma}{R} - 4\,\eta\,\frac{\dot{R}}{R} \rb.
\label{eq17}
\ee
At this point we note that to describe the Rayleigh collapse of the QGP clusters E.~Shuryak and P.~Staig \cite{21} used the following approximation form of (\ref{eq15})
\be
R\,\ddot{R} + \dfrac32\,{\dot{R}}^2 = - \frac{4\,\eta\,\dot{R}}{\rho\,R}.
\label{eq18}
\ee
The authors of \cite{21} came to the conclusion that the viscous dissipative flow may turn the collapse into a ``soft landing''. The discussion of this conclusion will be found below. 

\section{Bubble Dynamics in the Magnetic Field}

The quark matter formed in heavy ion collisions at RHIC and LHC is subjected to a strong magnetic field generated by spectator protons. At the initial stage the value of the field reaches $eB \simeq 10^{19}$ G \cite{45,46,47}. The time dependence of the created field is determined by the QGP electrical conductivity and by the evolution of the fireball geometry. We shall not discuss this complicated and controversial problem \cite{48,49,50,51,52}. 

To get insight into the bubble dynamics in the magnetic field we shall greatly oversimplify the problem and make several approximations. The external magnetic field is supposed to be constant and homogeneous. It will be assumed that the bubble always remains spherical. The fluid surrounding the bubble is assumed to be incompressible, irrotational ($\bf{v} = {\stackrel{\rightarrow}{\nabla}}\varphi$) and Newtonian (shear stress is proportional to the rate of shear strain). Dissipation is due to the two sources -- viscous losses and magnetic friction (see below). Magneto-hydrodynamics of such a liquid has been considered in \cite{41,53} and the expression for the magnetic force has been derived in \cite{41,54}. For the formalism developed in \cite{41,54} to be applicable the fluid needs to satisfy certain requirements in addition to the ones listed above. The fluid magnetic Mach number $R_M = 4 \pi \mu \sigma l v$ \cite{41,54} should be small, $R_M \ll 1$. Here $\mu$, $\sigma$, $l$ and $v$ are respectively the magnetic permeability, the electrical conductivity, the characteristic linear dimension and the characteristic velocity. For the bubble in the QGP the condition $R_M \ll 1$ is easily satisfied. The fluid magnetic permeability $\mu \sim 1$ (see below). The electrical conductivity has been calculated in a great number of papers, see, e.g., \cite{36,55,56,57,58,59}. A crude estimate $\sigma \sim 0.1 $ fm$^{-1}$ would suffice for our purposes since this research is aimed at the qualitative description of the phenomena. Worth mentioning that the value $\sigma \sim 0.1 $ fm$^{-1}$ is about five orders of magnitude larger than that for mercury or liquid copper. Nonetheless, the condition $R_M \ll 1$ may be fulfilled even for such a ``huge'' electrical conductivity. The bubble dimension is of the order $l = R \sim 1$ fm. Therefore $\sigma l \sim 10^{-1}$. According to \cite{21} the collapsing bubble may have a velocity about $4$ km/s, or $v \simeq 10^{-6}$ in natural units $c=1$. As a result, $R_M \ll 1$. 

We shall calculate the effect of the magnetic field on the bubble dynamics in two alternative ways. Both approaches will lead to the same result. First, we follow the reasoning provided in \cite{39,54}. The volume density of the force in magnetic field ${\bf f}$ (with dimension $M^5$) is given by a well-known formula \cite{39}
\be
\bf{f} = \left[\,\bf{j} \times \bf{B}\,\right].
\label{eq19}
\ee
The force ${\bf f}$ is due to the currents induced in the liquid moving with respect to the bubble. By virtue of the smallness of the magnetic Reynolds number the magnetic field in the current-currying region is assumed to be equal to the uniform and time-independent external one. Neglecting the perturbative expansion for ${\bf B}$ we may put $\Rot {\bf B} = \Div {\bf B} = 0$. The current in the moving liquid is
\be
\bf{j} = \sigma\,\left(\left[\,\bf{v} \times \bf{B}\,\right] + \bf{E} \right).
\label{eq20}
\ee
Th electric field $\bf{E}$ is potential since $\Rot \bf{E} = -\dfrac{\partial\bf{B}}{\partial t} = 0$. Then $\bf{E} = -\stackrel{\rightarrow}{\nabla}\varphi$ and the equation (\ref{eq20}) is replaced by
\be
\bf{j} = \sigma\,\left(\left[\,\bf{v} \times \bf{B}\,\right] - \stackrel{\rightarrow}{\nabla}\varphi \right).
\label{eq21}
\ee
Next we note that
\be
\Div\left[\,\bf{v} \times \bf{B}\,\right] = \bf{B}\left[\,\stackrel{\rightarrow}{\nabla} \times \bf{v}\,\right] - \bf{v}\left[\,\stackrel{\rightarrow}{\nabla} \times \bf{B}\,\right] = 0, 
\label{eq22}
\ee 
since the liquid is supposed to be irrotational and magnetic field is uniform. We also note that the term $\left[\,\bf{v} \times \bf{B}\,\right]$ has no component perpendicular to the bubble surface. It means that there is no induction of the electric charges on the bubble surface and $\Div \bf{j} = 0$. Equation (\ref{eq21}) with $\Div \left[\,\bf{v} \times \bf{B}\,\right] = \Div \bf{j} = 0$ yields $\Delta \phi = 0$ and the only contribution to the volume force stems from the first term in (\ref{eq21}):
\be
\bf{f} = \sigma\left[ [ \bf{v} \times \bf{B} ] \times \bf{B} \right] = - \sigma\left\{ \bf{v}\,\bf{B}^2 - \bf{B}\lb \bf{v}\bf{B} \rb\right\}.
\label{eq23}
\ee
Only the first ``drag'' term in (\ref{eq23}) contributes to the total force. The second term vanishes after integration over the polar angle with $\bf{B}$ taken as a polar axis.

Another way to calculate the force $\bf{f}$ is to consider the rate of the Joule energy dissipation \cite{39}. The Joule heating rate is caused by the currents induced in the liquid. The volume rate is equal to \cite{39}
\be
Q = \bf{j}^2/\sigma.
\label{eq24}
\ee
Resorting to the arguments presented above we conclude that only magnetic friction is responsible for the energy dissipation
\be
Q = \sigma [ \bf{v} \times \bf{B} ]^2.
\label{eq25}
\ee
The heat dissipation during the time interval $\delta t$ is $Q \delta t = -\bf{f} \bf{v} \delta t$, so that $Q = -\bf{f}\bf{v}$. Then
\be
\bf{f} = - \sigma\left\{ \bf{v}\,\bf{B}^2 - \bf{B}\lb \bf{v}\bf{B} \rb\right\}.
\label{eq26}
\ee
As expected, this result is identical to (\ref{eq23}). The presented derivation does not contradict the affirmation that Lorentz force does not produce work over the moving charge \cite{39}.

Now we can write the R-P equation with the magnetic field contribution included 
\be
R\,\ddot{R} + \dfrac32\,\dot{R}^{2} = \dfrac{1}{\rho}\left( p_i - p_0 - \frac{2\sigma}{R} - 4\,\eta\,\frac{\dot{R}}{R} - \sigma {\bf{B}}^2 R \dot{R} \right),
\label{eq27}
\ee
All terms in this equation are non-dimensional. One can ask a question \cite{25} of whether any term in the right-hand side of (\ref{eq21}) can halt the collapse. According to the scaling discussed in Sec.II the left-hand side of (\ref{eq21}) behaves as $(t_c-t)^{-6/5}$, the  surface term $\sigma/R \varpropto (t_c-t)^{-2/5}$, $\eta\dot{R}/R \varpropto (t_c-t)^{-1}$, $\sigma {\bf{B}}^2 R \dot{R} \varpropto (t_c-t)^{-1/5}$. Therefore it may be expected that none of these terms can prevent the bubble from the collapse. As stated in \cite{21} the above scaling may be violated by strong viscous dissipation (the term $4\eta\dot{R}/R$). The Rayleigh collapse turns into a ``soft landing'' if $\eta T/\rho \gtrsim 0.6$. Note that the shear viscosity is a pronounced function of the temperature \cite{60}. Close to $T_c$ $\eta/s$, where $s$ is the entropy density, takes its minimal value predicted to be $\eta/s = 1/(4\pi)$ \cite{61}. The physics at the last stages of the collapse is complicated. The equation of state of the media inside the bubble may play an important role. The sound radiation at $t \to t_c$ is very important since it diverges as $(t_c-t)^{-13/5}$ \cite{21,25}.

\section{The Itty-Bitty Magnetic Friction}

According to a common lore a huge magnetic field $eB \sim \Lambda^2_{QCD}$ created in heavy ion collisions brings substantial changes in physical observables like hadron masses, transport coefficients, etc \cite{32,33,34,35,36,37,59,62,63}. It turns out that this is not the case for the bubble dynamics. To see this let us first estimate the value of the magnetic force parameter $\sigma\,\bf{B}^2$. Predictions for the electrical conductivity both without  and with magnetic field vary in a wide range \cite{32,55,56,57,58,59,62,63}. The variance is really large as can be seen from Fig. 2 of \cite{63}. A reasonable estimate would be $\sigma \simeq 0.1$ fm$^{-1}$. As it was mentioned above this is a very high value as compared to the conductivity of the normal materials. What is the value of the magnetic field at the onset of a possible cavitation is a tough question. The time evolution of the magnetic field from its initial value $10^{19}$ G is a subject of discussions \cite{45,46,47,48,49,50,51,52}. We assume that magnetic field is diminished by 4-5 orders of magnitude from initial value, i.e., is equal to $10^{14}$ G. Then $\sigma\,{\bf{B}}^{2} \simeq 10^{-13}\text{ GeV}^{5}$. The ratio of the electromagnetic force to viscous force is characterized by the Hartmann numder \cite{39} $\operatorname{Ha} = (\sigma\,B^2\,l^2/\eta)^{1/2}$. The QGP shear viscosity depends on the temperature, density and magnetic field \cite{60,64,69}. Practically all calculations present the results in the form $\eta/s$, where $s$ is the entropy density. This expression is physically important as it shows that the QGP is the most ideal fluid in nature \cite{60,61}. The value of the shear viscosity itself $\eta \le 5 \cdot 10^{11}\text{Pa}\cdot\text{s}\simeq 0.1\text{ GeV}^3$ at $T=2\cdot10^{12}\text{K}\simeq 170$ MeV may be found in \cite{03}. It is interesting to estimate the value of $\eta$ corresponding to the critical value of the parameter $\eta\,T/\rho$ = 0.6 \cite{21}. Above this point the Rayleigh collapse turns into a ``soft landing''. Worth mentioning that the parametrization $\eta\,T/\rho \ge 0.6$ provides a correct dimension of $\eta$ but is misleading if one tries to deduce from it the temperature dependence of the shear viscosity \cite{60,71,72}. If we assume that cavitation takes place near $T_c$ we may take $T = 170$ MeV and $\rho \simeq (0.18-0.5)\text{ GeV}/\text{fm}^3$ \cite{03,70}. Such density is either slightly or a few times larger than the normal nuclear density. For definiteness we take $\rho = 0.3\text{ GeV}/\text{fm}^3$. The critical magnitude of $\eta$ corresponding to these values of $T$ and $\rho$ is $\eta \simeq 10^{-2}\text{ GeV}^3$. By the order of magnitude this is close to the results of \cite{73}. With $B=10^{14}\text{ G}$, $l=R=1\text{ fm}$, $\sigma=0.1\text{ fm}^{-1}$, $\eta=10^{-2}\text{ GeV}^3$ one gets 
\be
\operatorname{Ha} \simeq 1.6 \cdot 10^{-5}. 
\label{eq28}
\ee
Note that $(\operatorname{Ha})^2/4$ gives the ratio of magnetic and viscous terms in the R-P equation (\ref{eq27}).

\section{Summary and Discussion}

We came to the conclusion that an extremely strong magnetic field has only a tiny, not to say negligible, effect on the bubble dynamics in QGP. Its role is definitely small as compared to the viscous damping. In other words, the Joule magnetic dissipation is much smaller than the viscous one. The simple reason for this is that the QGP shear viscosity is very large albeit the QGP is the most ideal fluid in nature. As an example consider a bubble with a radius $R = 0.01$ cm in the sea water at $T=25^o$ C. The shear viscosity is $\eta = 8.9 \cdot 10^{-4}\text{ Pa}\cdot\text{s} \simeq 2\cdot 10^{-16}\text{ GeV}^{3}$, the electrical conductivity $\sigma = 5\text{ S}/\text{m} \simeq 1.5 \cdot 10^{-13}\text{ fm}^{-1}$ (numbers are from wiki). Then $H > 1$ already for $B > 10^7$ G.

Finally, we would like to rectify an important omission concerning the possible composition of matter inside and outside the bubble. Here we have to resort to frankly speculative considerations. The driving force for cavitation is the enhanced bulk viscosity \cite{08,09}. As it was shown in a number of works quark matter near $T_c$ possesses this property, see, e.g. \cite{74,75,76,77,78} and references therein. The role of a gas, or vapour, inside the bubble may be played by the hadron gas surrounded from the outside by quark matter. An interesting point in this case is that the magnetic permeability inside and outside will be different. The thermal QCD medium around the transition temperature is paramagnetic \cite{79,80} while the hadron gas inside may be weakly diamagnetic \cite{79}. 

\section*{Acknowlegments}
We would like to thank M.A.Andreichikov and Yu.A.Simonov for helpful discussions. We gratefully acknowledge the remarks of the Physical Review D Referee. This work was done in the frame of the scientific project supported by the Russian Science Foundation grant number 16-12-10414.




\end{document}